\documentclass[
aps,
pra,
superscriptaddress,
12pt,
reprint,
nofootinbib,
floatfix
]{revtex4-2}

\usepackage[utf8]{inputenc}
\usepackage[T1]{fontenc}
\usepackage[english]{babel}

\usepackage{graphicx}
\graphicspath{ {fig/} }

\usepackage{amsmath}
\usepackage{amssymb}
\usepackage{mathtools}

\usepackage[colorlinks=true, allcolors=blue]{hyperref}
\usepackage[all]{hypcap}  

\usepackage{braket}
\usepackage{xcolor}
\usepackage[normalem]{ulem}

\renewcommand{\vec}[1]{\boldsymbol{#1}}
\newcommand{\op}[1]{\hat{\vec{#1}}}
\newcommand{\V}{V}
\newcommand{\W}{W}
\newcommand{\D}{U}
\newcommand{\U}{S}

\begin{document}
\title{Vibronic interactions in trilobite and butterfly Rydberg molecules}
\date{\today}
\author{Frederic Hummel}
\email{hummel@pks.mpg.de}
\affiliation{Max-Planck-Institut für Physik komplexer Systeme, Nöthnitzer Str. 38, 01187 Dresden, Germany}
\author{Peter Schmelcher}
\email{pschmelc@physnet.uni-hamburg.de}
\affiliation{Zentrum für Optische Quantentechnologien, Fachbereich Physik, Universität Hamburg, Luruper Chaussee 149, 22761 Hamburg, Germany}
\affiliation{The Hamburg Centre for Ultrafast Imaging, Fachbereich Physik, Universität Hamburg, Luruper Chaussee 149, 22761 Hamburg, Germany}
\author{Matthew T. Eiles}
\email{meiles@pks.mpg.de}
\affiliation{Max-Planck-Institut für Physik komplexer Systeme, Nöthnitzer Str. 38, 01187 Dresden, Germany}

\begin{abstract}	
    Ultralong-range Rydberg molecules provide an exciting testbed for molecular physics at exaggerated scales.
    In the so-called \emph{trilobite} and \emph{butterfly} Rydberg molecules, the Born-Oppenheimer approximation can fail due to strong non-adiabatic couplings arising from the combination of radial oscillations and rapid energy variations in the adiabatic potential energy curves. 
    We utilize an accurate coupled-channel treatment of the vibronic system to observe the breakdown of Born-Oppenheimer physics, such as non-adiabatic trapping and decay of molecular states found near pronounced avoided crossings in the adiabatic potential curves.	
    Even for vibrational states localized far away from avoided crossings, a single channel model is quantitatively sufficient only after including the diagonal non-adiabatic corrections to the Born-Oppenheimer potentials.
    Our results indicate the importance of including non-adiabatic physics in the description of ultralong-range Rydberg molecules and in the interpretation of measured vibronic spectra. 
\end{abstract}

\maketitle

\section{Introduction}

Molecular states of Rydberg atoms and atoms or ions feature exotic binding mechanisms and offer novel opportunities to investigate fundamental physics.
In ultralong-range Rydberg molecules, the Rydberg atom binds to ground-state atoms via an attractive interaction induced by the scattering of the Rydberg electron from the atoms \cite{Greene2000,Bendkowsky2009,Fey2019rev}.
In charged Rydberg molecules, the Rydberg atom binds to an ion due to long-range electrostatic interactions leading to an avoided crossing of an attractive polarization potential and a repulsive dipole potential \cite{Deiss2021,Duspayev2021,Zuber2022}. 
In Rydberg macrodimers, two Rydberg atoms bind at the avoided crossing of attractive and repulsive van-der-Wals potentials \cite{Boisseau2002,Overstreet2009,Sassmannshausen2016}.
All of these molecules feature bond lengths on the 0.1-10~$\mu$m scale leading to slow vibrational dynamics which can be observed in real time \cite{Zou2022}.

Due to the high density of states in Rydberg atoms and the correspondingly small energy gaps, non-adiabatic couplings play a crucial role in Rydberg molecules.
They are known to modify the vibrational spectrum of macrodimers \cite{Hollerith2019} and provide a decay channel for charged Rydberg molecules \cite{Duspayev2022}.
In ultralong-range Rydberg molecules, non-adiabatic couplings have been considered  at the level of the Landau-Zener approximation to predict electronic transitions \cite{Schlagmuller2016x}.
Conical intersections in their molecular potentials can be exploited to control atomic collisions and lifetimes over a wide range of principle quantum numbers and temperatures \cite{Hummel2021prl}.

Ultralong-range Rydberg molecules have very unusual properties, such as oscillatory Born-Oppenheimer potentials, rapid changes of electronic character, and extreme sensitivity to external fields.
Therefore, they provide a unique environment to test molecular physics on slow vibrational time scales, allowing direct observation of effects arising from the entanglement of vibrational and electronic degrees of freedom and offerering abundant opportunities to examine concepts of molecular physics in energetic and spatial regimes very different from those relevant for ground state molecules.
Despite these promising opportunities, an ab initio treatment of non-adiabatic couplings in ultralong-range Rydberg molecules has not been performed yet.

In this article, we study ultralong-range Rydberg molecules in a complete  coupled-channel treatment for the electronic and vibrational degrees of freedom.
The arising non-adiabatic couplings lead to vibronic interactions. 
We focus on molecular states formed in the vicinity of the avoided crossing of the trilobite and butterfly electronic states. 
These polar molecules have high electronic orbital angular momentum character and their non-adiabatic coupling can become singular.
We discuss several strategies for diabatization of the adiabatic representation of the total molecular wave function in order to solve the molecular Hamiltonian.

We find that non-adiabatic corrections to the molecular potentials are crucial for an accurate interpretation of molecular spectra.
This sheds new light on measurements of electron scattering phase shift relying on Rydberg molecules \cite{Anderson2014exp,Sassmannshausen2015,Boettcher2016,MacLennan2019,Engel2019}, as the level shifts caused by non-adiabatic corrections could be mis-identified as stemming from inaccurate scattering properties.
Furthermore, off-diagonal non-adiabatic couplings lead to the formation of molecular states without single-channel counterparts. 
Their structure resembles the quantum reflection states observed in single channel calculations \cite{Bendkowsky2010,Anderson2014th}, although they are classically trapped by a potential barrier created by non-adiabatic coupling.
Other classes of molecular states decay non-adiabatically.
Spectral signatures for the experimental observation of the vibronic interaction effects are discussed.
	
This manuscript is structured as follows. 
In section \ref{sec:theory}, we lay out the theory and methods used throughout.
Section \ref{sec:non-adibatic} introduces in detail the non-adiabatic representation before we elaborate on diabatization in \ref{sec:diabatic}.
This is to promote and simplify adaptation of this framework for future investigations.
Details of the model for ultralong-range molecules and the physical parameters are provided in \ref{sec:parameters}.
Section \ref{sec:results} contains the results, firstly introducing the adiabatic potential energy curves in \ref{sec:adiabatic}, secondly considering variations in the electronic structure in \ref{sec:electronic}, thirdly providing the molecular spectra in \ref{sec:vibrational}, and finally focusing on vibronic interactions in \ref{sec:off-diagonal}.
We provide a conclusion and an outlook in section \ref{sec:conclusion}.

\section{Theory \& Methods \label{sec:theory}}

We consider two atoms, one excited to a high-lying Rydberg state, separated by a large distance $R$ along the $z$ axis of the coordinate system. 
The total Hamiltonian reads (in atomic units)
\begin{equation}
	\hat{H}=-\frac{\vec{\nabla}_R^2}{2\mu}-\frac{\vec{\nabla}_r^2}{2}+V_c(\vec{r})+V_{ea}(\vec{r},R), \label{eq:ham}
\end{equation}
where $\mu$ is the reduced mass of the diatomic system, $\vec{r}$ is the coordinate of the Rydberg electron, $V_c$ is the interaction between the Rydberg electron and the ionic core, and $V_{ea}$ the interaction of the Rydberg electron with the distant ground-state atom. 
The last three terms of equation (\ref{eq:ham}) form the electronic Hamiltonian $\hat{H}_e(\vec{r},R)$ consisting of the corresponding kinetic energy and two potential terms, one for each heavy particle.
To solve $\hat{H}$, we employ the \emph{adiabatic representation} and expand the total wave function in terms of solutions of the electronic Hamiltonian \cite{Born1927, Born1954}
\begin{equation}
    \Psi(\vec{r},R)=\sum_\alpha \chi_\alpha(R) \psi_\alpha(\vec{r};R), \label{eq:totalwave}
\end{equation}
such that the sum over $R$-dependent coefficients $\sum_\alpha|\chi_\alpha|^2$ can be interpreted as a vibrational wave function of the nuclei, and 
\begin{equation}
    \hat{H}_e(\vec{r},R) \psi_\alpha(\vec{r};R)=\V_\alpha(R) \psi_\alpha(\vec{r};R),
\end{equation}
defines the adiabatic electronic eigenstates $\psi_\alpha$ with eigenvalue $\V_\alpha$ given a fixed nuclear configuration $R$. 
$\V_\alpha(R)$ are the adiabatic potential energy curves.

\subsection{Non-adiabatic couplings \label{sec:non-adibatic}}

Applying the Hamiltonian (\ref{eq:ham}) to equation (\ref{eq:totalwave}) and projecting out the electronic eigenstates leads to a general set of coupled-channel, radial Schrödinger equations for the vibrational motion \cite{Koppel1984}
\begin{equation}
	\left[ -\frac{\nabla_R^2}{2\mu}+\V_\alpha(R)-E\right] \chi_\alpha(R)=\frac{1}{2\mu}\sum_{\alpha'} \Lambda_{\alpha \alpha'} \chi_{\alpha'}(R), \label{eq:SE}
\end{equation}
for a total energy $E$.
$\nabla_R$ indicates the derivative with respect to the radial, nuclear coordinate $R$.
The right-hand side contains the non-adiabatic coupling operator
\begin{equation}
	\Lambda_{\alpha \alpha'} = 2 P_{\alpha\alpha'} \nabla_R+Q_{\alpha\alpha'},
\end{equation}
which consists of the derivative coupling elements
\begin{equation}
	P_{\alpha\alpha'} = \int \mathrm{d} \vec{r} \, \psi^\ast_\alpha(\vec{r};R) \nabla_R \psi_{\alpha'}(\vec{r};R),
\end{equation}
and the non-adiabatic scalar coupling elements
\begin{equation}
	Q_{\alpha\alpha'} = \int \mathrm{d} \vec{r} \, \psi^\ast_\alpha(\vec{r};R) \nabla^2_R \psi_{\alpha'}(\vec{r};R).
\end{equation}
For all practical purposes, the sum in equations (\ref{eq:totalwave}) and (\ref{eq:SE}) needs to be truncated.
In the \emph{Born-Oppenheimer approximation}, all non-adiabatic terms are neglected.
The right-hand side of equation (\ref{eq:SE}) vanishes and the resulting decoupled equations can be solved individually. 
In the \emph{Born-Huang approximation}, only the diagonal non-adiabatic terms are considered. 
In many cases, these terms are crucial to capture qualitative corrections \cite{Greene2017,Naidon2017}.
It is helpful to absorb the diagonal couplings in the potential energy curves through the definition
\begin{equation}
    \W_\alpha(R) = \V_\alpha(R) - \frac{1}{2\mu}Q_{\alpha\alpha}(R).
\end{equation}
Note that $P_{\alpha\alpha}=0$, due to anti-hermiticity of the derivative coupling operator.
Since the diagonal elements $Q_{\alpha\alpha}$ are strictly negative (compare equation (\ref{eq:P2}) below), this leads to a shift to larger potential energies $\W_\alpha\geq\V_\alpha$. 

In principle, the non-adiabatic coupling elements are straightforwardly obtained from the adiabatic electronic states.
However, numerical difficulties frequently arise in practice. 
With respect to the electronic states, the operator $\Lambda_{\alpha\alpha'}$ is hermitian, while $P_{\alpha\alpha'}$ is anti-hermitian. 
Hence, $Q_{\alpha\alpha'}$ is non-hermitian and numerical inaccuracies or round-off errors in its computation can easily lead to an overall non-hermitian matrix and non-physical complex energy eigenvalues. 
This can be avoided by recasting
\begin{equation}
	Q_{\alpha\alpha'} = P^2_{\alpha\alpha'} + \nabla_R P_{\alpha\alpha'}. \label{eq:PQ}
\end{equation}
and using partial integration to ensure hermiticity of the matrix elements of the non-adiabatic coupling.
To this end, let us introduce a basis of the vibrational Hilbert space $B_i(R)$, such that any $R$-dependent function can be conveniently expanded as
\begin{equation}
	f(R) = \sum_i c_i B_i(R),
\end{equation}
with constant coefficients $c_i$.
The matrix elements of the non-adiabatic coupling operator in this basis are
\begin{align}
	\nonumber & \int \mathrm{d} R \, B_i B_j \Lambda_{\alpha\alpha'} \\
	& \quad = \int \mathrm{d} R \, [(B_i B'_j - B'_i B_j) P_{\alpha\alpha'} + B_i B_j P^2_{\alpha\alpha'}].
\end{align}
As the first term is the product of two anti-hermitian operators in the vibrational and electronic Hilbert sub-spaces, respectively, and the second term is the product of two hermitian operators, this guarantees a total hermitian representation.
It additionally avoids the need to compute second derivatives of the electronic wave functions, because
\begin{equation}
	P^2_{\alpha\alpha'} = -\int \mathrm{d} \vec{r} \, \nabla_R \psi^\ast_\alpha(\vec{r};R) \nabla_R \psi_{\alpha'}(\vec{r};R). \label{eq:P2}
\end{equation}

In conclusion, we can solve equation (\ref{eq:SE}), e.g.~by performing an exact diagonalization in the electronic and the vibrational degree of freedom. 
In some circumstances, however, the derivative coupling elements may contain highly singular points such that equation (\ref{eq:SE}) becomes formally  ill-defined \cite{Worth2004}. 
It is then reasonable to switch to a representation that circumvents such difficulties. 

\subsection{Diabatic representation \label{sec:diabatic}}

We can recast equation (\ref{eq:SE}) in the following form
\begin{equation}
	\left[ -\frac{1}{2\mu}(\nabla_R+\op{P})^2+\op{\V}(R)-E\right] \vec{\chi}(R)=0, \label{eq:trSE}
\end{equation}
where the first term is understood to contain a dressed kinetic energy operator that has absorbed all non-adiabatic couplings. 
Note that equation (\ref{eq:trSE}) is a matrix equation where $\op{\V}$ is a matrix that contains the $\V_\alpha$ along the diagonal and $\vec{\chi}$ is a vector of all $\chi_\alpha$. 

Equation (\ref{eq:trSE}) is invariant under unitary transformations $\op{\U}(R)$.
Exact diabatization is achieved when $\op{\U}$ satisfies \cite{Baer2006}
\begin{equation}
	(\nabla_R \op{\U} + \op{P} \op{\U})=0,\label{eq:trans}
\end{equation} 
in which case equation (\ref{eq:trSE}) becomes
\begin{equation}
	\left[ \frac{-\nabla_R^2}{2\mu}+\op{\D}(R)-E\right] \tilde{\vec{\chi}}(R)=0, \label{eq:diaSE}
\end{equation}
where $\tilde{\vec{\chi}}=\op{\U}^\dagger \vec{\chi}$ is the total wave function expressed in the diabatic representation and $\op{\D}=\op{\U}^\dagger \op{\V} \op{\U}$ is the diabatic potential matrix. 
In general, $\op{\U}$ is not unique, but as long as there is only a single adiabatic parameter (in our case $R$), equation (\ref{eq:trans}) does possess a unique solution and can be solved analytically \cite{Top1975}. 
Considering two adiabatic electronic states, $\op{\U}$ is given as a simple rotation about the angle
\begin{equation}
	\vartheta(R) = \int_R^\infty \mathrm{d} R' \, P_{12}(R').
\end{equation}
The solution for three and four electronic states is given in Appendix \ref{sec:transformation}.
There exist other strategies for diabatization \cite{Smith1969, Pacher1988, Alekseyev2000}, typically related to achieving an electronic representation that provides continuity of some physical observable, such as electronic dipole moments or charge distributions. These are especially useful if $\op{\U}$ is not known. We discuss dipole diabatization and a related approach in Appendices \ref{sec:dipole} and \ref{sec:hybrid}.

\subsection{Ultralong-range Rydberg molecules \label{sec:parameters}}

We consider ultralong-range Rydberg molecules of rubidium.
The adiabatic potential energy curves are obtained from diagonalizing the electronic Hamiltonian in a basis of atomic states $\phi_{nlm}$. 
These states are labeled by the principle and orbital angular momentum quantum numbers of the Rydberg atom $n$, $l$, and $m_l$, and satisfy the Rydberg atom's Schrödinger equation 
\begin{equation}
	\left[-\frac{\vec{\nabla}_r^2}{2}+V_c(\vec{r})\right]\phi_{nlm}(\vec r) = E_{nl}\phi_{nlm}(\vec r),
\end{equation}
where $E_{nl}=-\frac{1}{2(n-\mu_l)^2}$ and $\mu_l$ is an angular-momentum-dependent, but energy-independent quantum defect.
These quantum defects detune the atomic states with low angular momentum  $l<3$ from the degenerate Rydberg manifold $-\frac{1}{2n^2}$. 
We use quantum defect data from \cite{Li2003}. 

The electron-atom interaction is modeled using the Fermi pseudopotential \cite{Fermi1934,Omont1977}
\begin{align}
	\nonumber V_{ea}(\vec{r},R) = \ &2\pi a_s(R) \delta(\vec{r}-R\vec{e}_z) \\
	& +6\pi a_p(R) \vec{\nabla}_{\vec{r}} \delta(\vec{r}-R\vec{e}_z) \vec{\nabla}_{\vec{r}}, \label{eq:int}
\end{align}
where $a_{s(p)}$ is the scattering length (volume) in the $s$-($p$)-wave channel. 
These depend on $R$ via the semiclassical kinetic energy of the electron, $k^2=2E_{nl}+2/R$. 
We use scattering data from \cite{Engel2019}, where the zero-energy $s$-wave scattering length is $a_s(0)=-15.24 \, a_0$.

To represent the electronic degrees of freedom for a Rydberg molecule formed by a Rydberg atom excited to the principal quantum number $n$, we include in our atomic basis all states with $n$ and $n-1$. 
From this, we obtain good qualitative agreement with more accurate Kirchhoff integral methods based on the Coulomb Green's function \cite{Hamilton2002, Chibisov2002}.
This descriptive level serves our goal to quantify and elucidate the importance of non-adiabatic physics, which can be obscured by the complexity of the full molecular picture.
This would include quantitative corrections to the electronic structure resulting from higher partial wave scattering \cite{Giannakeas2020b} and the electronic and nuclear spins of both atoms \cite{Khuskivadze2002, Eiles2017}. To represent the vibrational degree of freedom, we employ B splines \cite{Bachau2001}. 
Due to the banded structure of the involved matrices, this is numerically very efficient.
Depending on the principle quantum number, we use 500 to 1500 splines of order 12 and a constant spacing of knot points to converge the vibrational energies and wave functions.
	
To simulate an absorption spectrum, the standard experimental observable to study Rydberg molecules, we account for the dipole transition strength of the total wave function assuming a two-photon excitation from the $5s$ ground state via the $6p$ intermediate state and for the Franck-Condon factor assuming a homogeneous and isotropic gas of ground-state atoms, i.e.~a uniform distribution of the initial vibrational state:
\begin{equation}
	\sigma(E) \propto \int \mathrm{d} R \, R^2 \int \mathrm{d} \vec{r} \, \vec{r} \phi_{6p}(\vec{r}) \Psi(\vec{r},R),
\end{equation}
where the wave functions are defined in equation (\ref{eq:totalwave}) and below.
Specifically, only the atomic $S$- and $D$-state character of the total wave function contribute to the spectral signal. 
Naturally, a one-photon transition directly from the $5s$ ground state changes the spectral characteristics.
Furthermore, a Gaussian profile for the line broadening of the excitation laser with a width of 1~MHz is assumed.

\section{Results \& Discussion \label{sec:results}}

In the following, we present the characteristic properties of the electronic, the vibrational, and the spectral structure of neutral Rydberg molecules taking into account non-adiabatic couplings. 
To illustrate the main physical phenomena, we focus on a representative subset of principle quantum numbers $n=43$ to 45, through which we can elaborate on generic $n$-dependent features that we found studying a larger range between $n=25$ and 70.

\subsection{Structure of the adiabatic potential curves \label{sec:adiabatic}}

\begin{figure}
	\centering
	\includegraphics[width=0.4\textwidth]{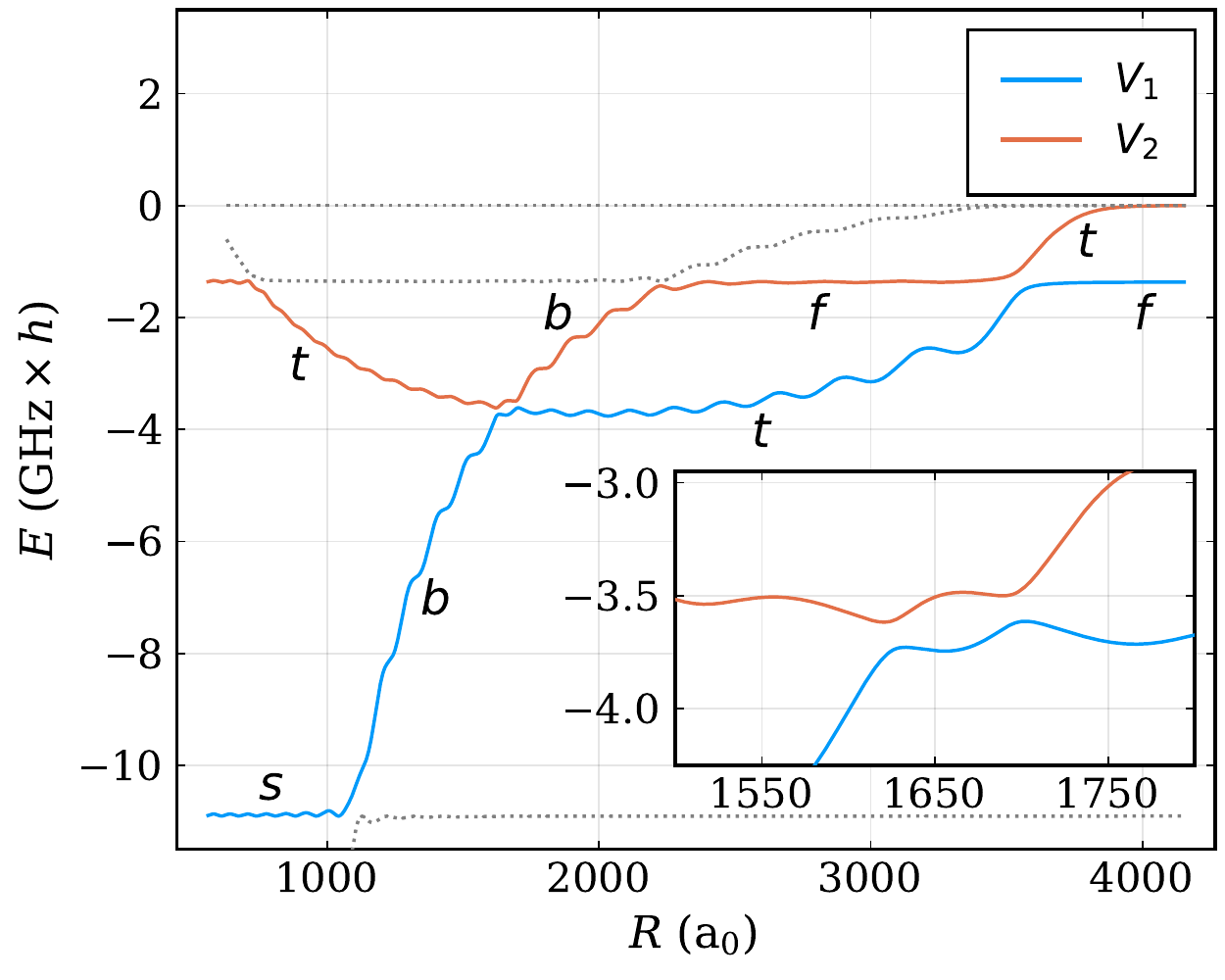}
	\caption[Adiabatic potentials for $n=43$]{Adiabatic potential energy curves $\V_1$ and $\V_2$ for principle quantum number $n=43$. Other adiabatic potentials are printed as gray, dotted lines. The inset shows a zoom of the avoided crossings. Additional labels show the dominant electronic character of these adiabatic states, which varies with $R$: the atomic $S$ state (s), trilobite (t), butterfly (b), and $F$ state (f). Energies are given relative to the degenerate atomic Rydberg level. }
	\label{fig:pot43}
\end{figure}

Let us first consider the adiabatic potential curves shown in figure \ref{fig:pot43} for $n=43$.
The level structure of Rydberg molecules is such that these fall into two classes.
The electronic states of dominant low-$l$ character are, due to their quantum defects, only weakly perturbed by the interaction with the ground state atom.
Their potential curves are, on the scale of figure \ref{fig:pot43}, nearly flat.
Two additional adiabatic potential energy curves split away from the degenerate Rydberg manifold.
Loosely speaking, each of these can be associated with the interaction channels given by the two terms in equation (\ref{eq:int}).
The resulting $s$-wave dominated and $p$-wave dominated molecular states are called trilobite and butterfly states, respectively \cite{Greene2000,Hamilton2002}.
These two states have a prominent avoided crossing \cite{Chibisov2002} shown in the inset.
Non-adiabatic effects are expected to be strongest in its vicinity. 

Based on this Rydberg level structure, the adiabatic eigenstates $\psi_1$ and $\psi_2$ defining the potentials $\V_1$ and $\V_2$ shown in figure \ref{fig:pot43} are subject to several changes in the dominant electronic character. 
Three avoided crossings occur for $V_1$ (solid blue). 
Starting asymptotically at large $R$, $V_1$ has dominant $F$-state character. 
Going inward towards smaller $R$-values, firstly $V_1$ attains trilobite character, secondly attains butterfly character, and lastly, at much lower energies, attains $S$-state character. 
For $V_2$ on the other hand (solid red) four avoided crossings occur. 
Starting asymptotically at large $R$, $V_2$ has dominant trilobite character. 
Going inward towards smaller $R$-values, firstly $V_2$ attains $F$-state character, secondly attains butterfly character, thirdly attains trilobite character, and lastly attains $F$-state character once more. 

We obtain converged results for the vibronic spectra in the examined energy ranges shown in the following sections by selecting only these two adiabatic states for the calculation. 
No deviations were found between these calculations and those performed with three and four channels, including also the asymptotic $S$ and $F$ adiabatic potentials shown in figure \ref{fig:pot43}.
Additionally, to obtain molecular states, we set a variable hard wall boundary at small $R_0$ and average the results over 50 different $R_0$ values.
This is the stabilization method established for ultralong-range molecules in reference \cite{Bendkowsky2010}.

\subsection{Variation of the electronic structure \label{sec:electronic}}

\begin{figure*}
	\centering
	\includegraphics[width=\textwidth]{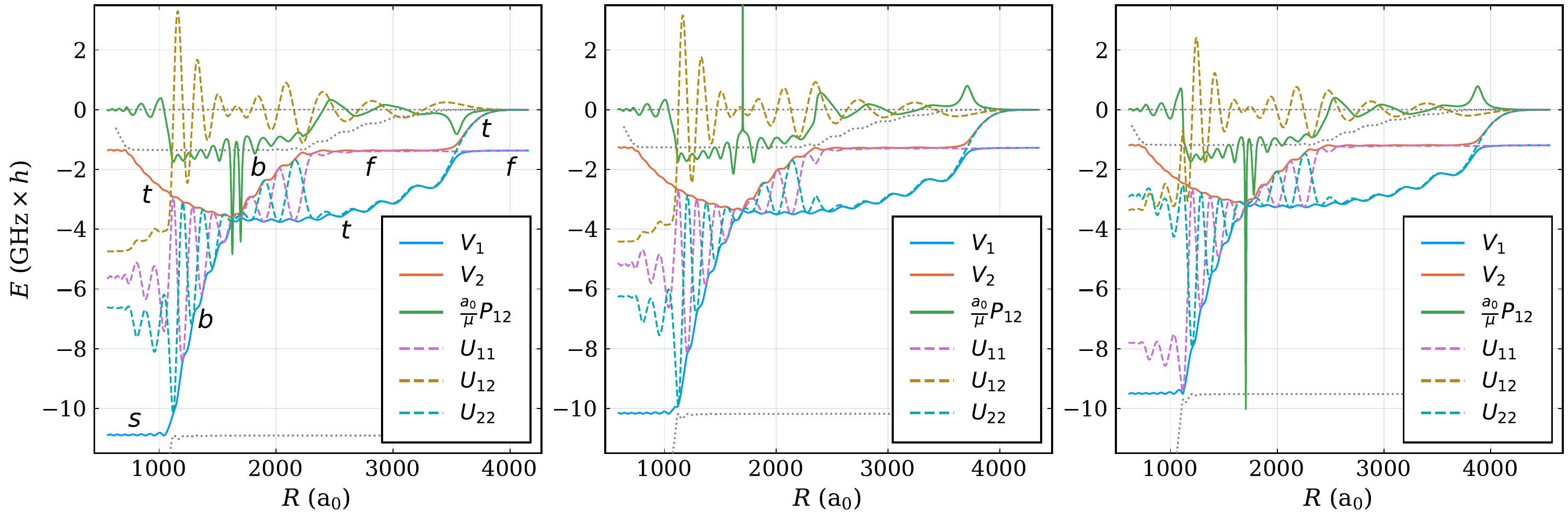}
	\caption[Adiabatic and diabatic potentials for $n=43$ to 45]{Adiabatic potential energy curves $\V_1$, $V_2$, the off-diagonal derivative coupling element $P_{12}$, and the diabatic potential matrix elements $\D_{11}$, $\D_{12}$, and $\D_{22}$ for principle quantum numbers $n=43$ to 45 (left to right). Other adiabatic potentials are printed as gray, dotted lines, additional labels as in figure \ref{fig:pot43}. Energies are given relative to the degenerate atomic Rydberg level. }
	\label{fig:pot}
\end{figure*}

Let us consider how the adiabatic potential energy curves, the off-diagonal derivative coupling element, and the diabatic potential curves, vary as a function of $n$. 
These quantities are plotted in figures \ref{fig:pot}. 
The derivative coupling element $P_{\alpha\alpha'}$ is a prefactor to a derivative operator of the vibrational degree of freedom in the total Schrödinger equation (\ref{eq:SE}). 
Therefore, multiplying $P_{\alpha\alpha'}$ by a unit length and dividing by the molecule's reduced mass results in a unit of energy, allowing us to show it alongside the electronic adiabatic potential energy curves.
Hence, we can compare the relative strengths of non-adiabatic couplings for molecules at different $n$. 
	
All of the avoided crossings discussed previously are accompanied by signatures in the derivative coupling element $P_{12}$, shown as solid green line.
Note for example the peak at the outermost avoided crossing at large $R$.
The most unusual feature of $P_{12}$ is its relatively large value over the range where the two adiabatic states have trilobite and butterfly character at the same $R$ values.
This is also the region where $P_{12}$ has the most significant feature, namely a one or two (depending on the principle quantum number) sharp peaks around the avoided crossing.
At $n=43$, in the left-most subfigure, we find a symmetric double peak structure in $P_{12}$.
At $n=44$, in the middle subfigure, we find a very sharp peak in $P_{12}$ that extends beyond the range of the figure.
At at $n=45$, in the right-most subfigure, we find an asymmetric double peak structure in $P_{12}$.
These are the main characteristics of the derivative coupling element for the two underlying adiabatic states over the full range of $n$ between 25 and 70.
The patterns repeat in increments of approximately 3 to 4 in $n$. 
In the case of a single sharp peak, the avoided crossing of the two potentials becomes extremely narrow.
This is remarkable given the overall energies of the molecular system.
It resembles the cases of conical intersections discussed in reference \cite{Hummel2021prl}.
Due to the singular character of the coupling elements, methods to solve the total Hamiltonian (\ref{eq:ham}) that rely on the adiabatic representation fail.
It is useful to transform to the diabatic representation.

The elements of the diabatic potential matrix are shown as dashed lines in figure \ref{fig:pot}. 
The diagonal elements (diabatic potential curves) are shown in purple and blue and the off-diagonal coupling is depicted in beige.
Notably, over all regions where the dominant character of the adiabatic states are trilobite and butterfly, all elements of the diabatic potential matrix have strong oscillatory character.
As the rotation angle of the transformation between the adiabatic and diabatic representation is the integral of the derivative coupling element $P_{12}$, this behavior may not seem surprising.
After all, $P_{12}$ is large in that entire region.
It is, however, rather unusual, as the diabatic potentials do not provide any physical intuition for the molecular system, as would be desirable for diabatic potentials.
The oscillatory behavior is known, however, to occur for states which are coupled by a large dipole transition element \cite{Esry2003}.
This is certainly true for the trilobite and butterfly state. 
We notice that the $n$-dependent structure of the derivative coupling element does not have a significant influence on the structure of the diabatic potential matrix elements. 
A simplified model including only trilobite and butterfly and neglecting quantum defect states is studied in Appendix \ref{sec:quantum-defects} reproducing these diabatic potentials.

\begin{figure}
	\centering
	\includegraphics[width=0.4\textwidth]{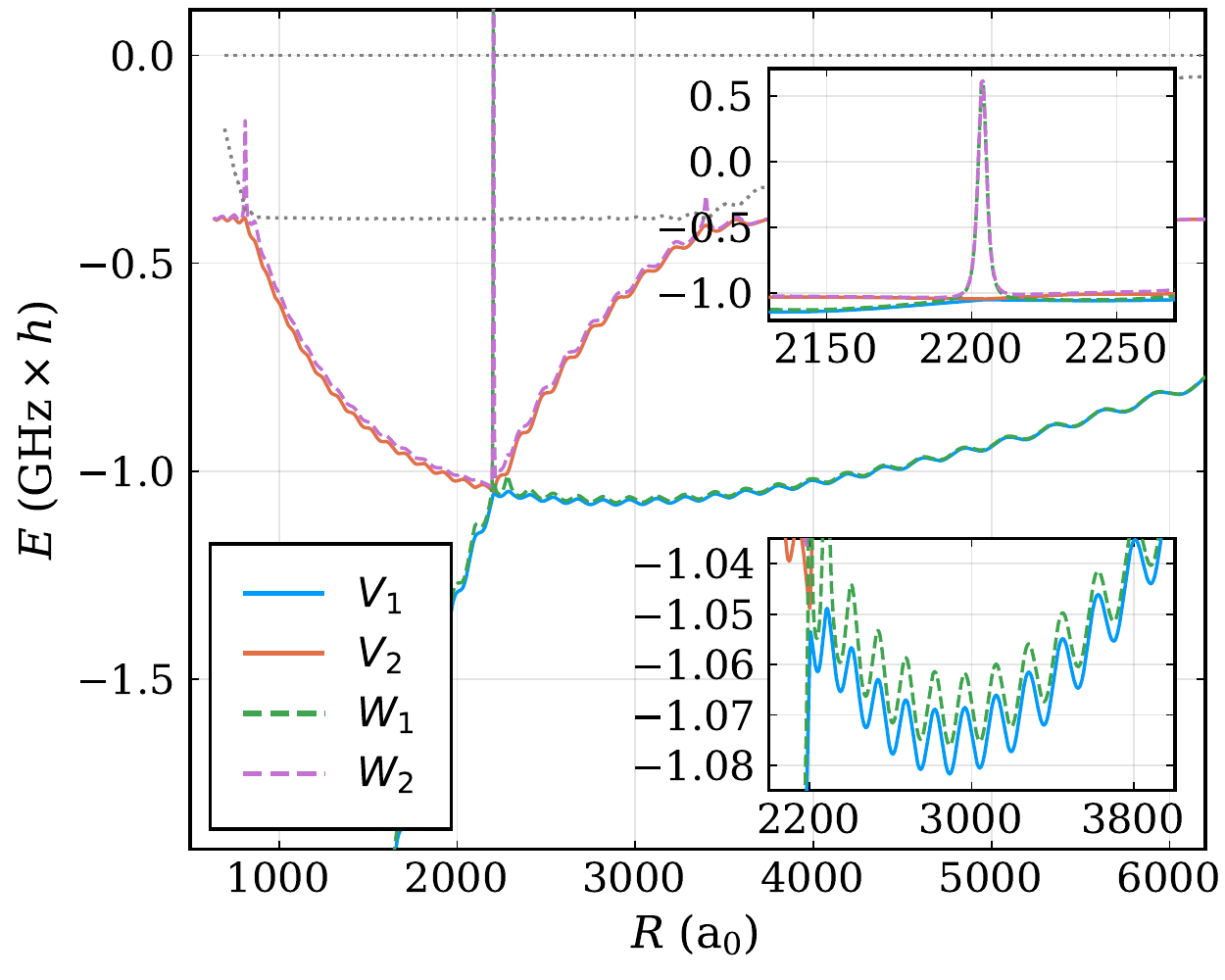}
	\caption[Adiabatic and Born-Huang potentials for $n=65$]{Adiabatic potential energy curves $\V_1$, $\V_2$, and Born-Huang corrected potentials $\W_1$, $\W_2$ for principle quantum number $n=65$. The insets show zooms of relevant regions. Energies are given relative to the degenerate atomic Rydberg level.}
	\label{fig:pot65}
\end{figure}

The overall magnitude of the non-adiabatic coupling elements, both for $P_{\alpha\alpha'}$ and $Q_{\alpha\alpha'}$, is approximately constant across all $n$. 
On the other hand, the energies of the adiabatic potentials scale as $n^{-3}$, in line with the scaling of the level splittings in Rydberg atoms.
As a result, non-adiabatic effects play a more important role at large $n$.
The diagonal scalar coupling terms lead to an overall energy shift of the potentials on the order of 5~MHz. 
At low $n\lesssim30$, this is more or less negligible. 
However, at large $n\gtrsim60$, this shift is larger than the energy spacing of molecular states. 
Figure \ref{fig:pot65} shows the adiabatic potentials for $n=65$ as solid lines along with the Born-Huang corrected potentials, which include the diagonal scalar coupling, as dashed lines. 
The lower inset shows that the Born-Huang terms give rise to a correction on the order of 5~MHz, which leads to a significant overall shift of the molecular states. 
Additionally, the upper inset shows the potential barrier that appears at the narrow avoided crossing due to these diagonal non-adiabatic coupling terms. This barrier is a result of the narrow energy gap of the two states at the crossing and the correspondingly sharply peaked derivative coupling element.

\subsection{Vibrational structure and molecular spectrum \label{sec:vibrational}}

\begin{figure*}
	\centering
	\includegraphics[width=\textwidth]{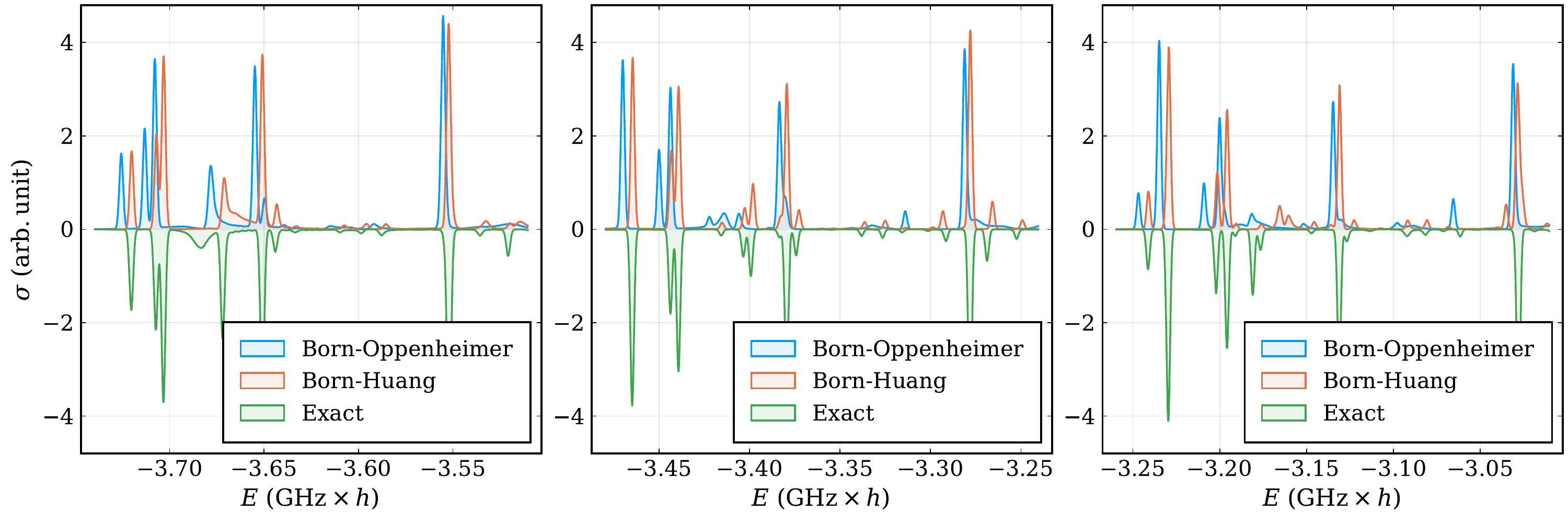}
	\caption[Simulated absorption spectra for $n=43$ to 45]{Simulated absorption spectra employing a single channel neglecting (including) scalar coupling, i.e.~using Born-Oppenheimer (Born-Huang) potentials, and employing coupled channels for principle quantum numbers $n=43$ to 45 (left to right).  Energies are given relative to the degenerate atomic Rydberg level.}
	\label{fig:spec}
\end{figure*}

Let us now turn to the energy eigenfunctions of the total Hamiltonian (\ref{eq:ham}).
Figure \ref{fig:spec} shows the simulated spectral signal obtained from the total wave functions depending on the underlying model:
In blue, the single channel model using the adiabatic potential energy curves $\V_1$ and $\V_2$, the Born-Oppenheimer spectrum.
In orange, the single channel model using the Born-Huang corrected potential energy curves $\W_1$ and $\W_2$, which include the diagonal scalar couplings $Q_{11}$ and $Q_{22}$, the Born-Huang spectrum.
And in green (mirrored for visibility), the two-channel model in the diabatic representation, the exact spectrum.

Two features are immediately striking. 
Firstly, both the Born-Huang spectrum and the exact spectrum are shifted in energy compared to the Born-Oppenheimer spectrum.
Secondly, the Born-Huang spectrum reproduces the exact spectrum quite accurately over most of the energy range. 
This clearly shows the importance of including at least the diagonal non-adiabatic couplings for investigations of ultralong-range Rydberg molecules. 
That said, we do find signals in the exact spectrum that have no counterpart in the single channel spectra.
At $n=43$, in the left-most subfigure, around $E=-3.68\,$GHz there is a broad peak next to a sharp peak with rather strong signal. 
The single channel spectra only show a single asymmetric peak of weaker signal.
Around $E=-3.51\,$GHz, there is a weak peak in the exact signal not found in the single channel spectra.
At $n=44$, in the middle subfigure, the Born-Huang spectrum is astonishingly accurate. 
Some of the peaks with weaker signal are slightly shifted towards higher energies compared to the exact spectrum.
At $n=45$, in the right-most subfigure, around $E=-3.18\,$GHz the exact spectrum features a double peak structure significantly shifted from a similar but weaker in signal double peak in the Born-Huang spectrum.
Interestingly, the Born-Oppenheimer spectrum predicts a single peak at the correct energy.
Between $E=-3.03$ and $E=-3.07\,$GHz, the single channel spectra each predict a peak completely absent in the exact spectrum.

\subsection{Off-diagonal non-adiabatic couplings \label{sec:off-diagonal}}

\begin{figure*}
	\centering
	\includegraphics[width=\textwidth]{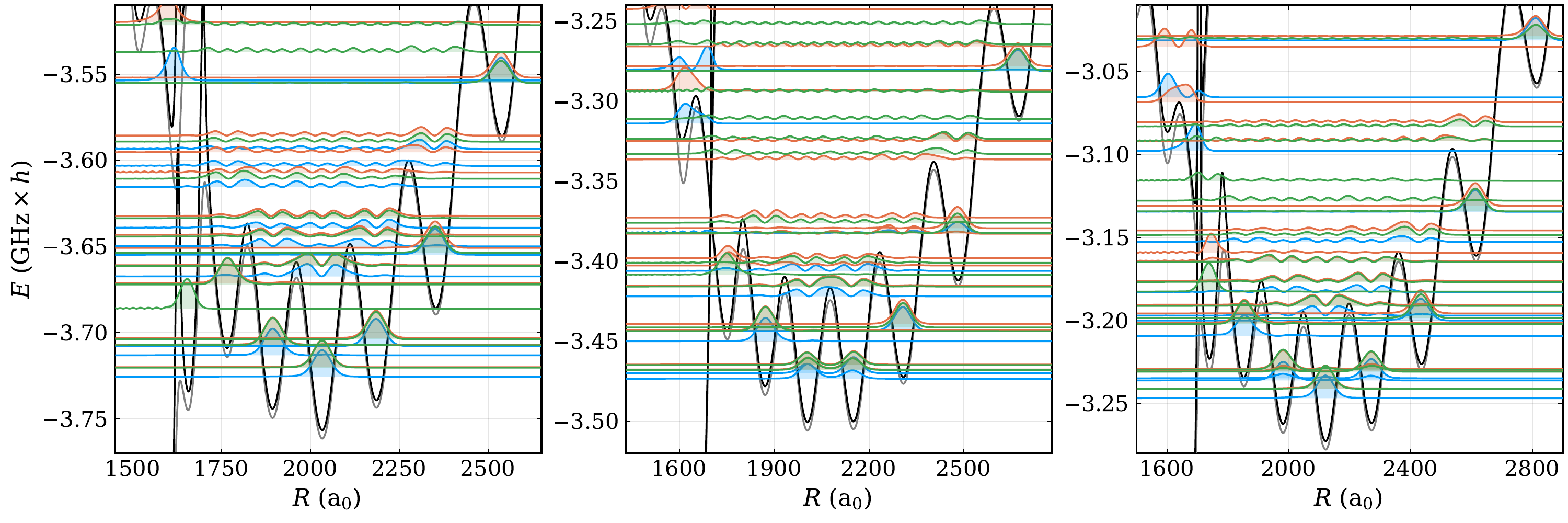}
	\caption[Molecular eigenstates for $n=43$ to 45]{Selected molecular eigenstates obtained using a single channel calculation excluding (blue) and including (orange) scalar coupling as well as those from the two coupled channels calculation (green). The Born-Oppenheimer (gray) and Born-Huang (black) potentials for principle quantum numbers $n=43$ to 45 (left to right) are also shown. These states correspond to the spectral signals shown in figure \ref{fig:spec}. Energies are given relative to the degenerate atomic Rydberg level.}
	\label{fig:wave}
\end{figure*}

Some of these spectral features have their origin in strong off-diagonal non-adiabatic couplings that cannot be described in a single channel model.
In order to gain a better understanding of these situations, we now turn to study the molecular wave functions directly.
Figure \ref{fig:wave} shows selected molecular eigenstates after tracing over the electronic degree of freedom, shifted according to their eigenenergies.
The corresponding squared wave functions can therefore be interpreted as vibrational probability distributions.
Here, states that are strongly suppressed in the spectra due to their vanishing Franck-Condon overlaps are not shown. 
This holds also for states resembling box states, which are artifacts related to the inner hard-wall boundary.
The energy window is the same as that for the spectra plotted in figure \ref{fig:spec}. 
The gray curves represent the adiabatic potentials $\V_1$ and $\V_2$ shown in figure \ref{fig:pot}.
Additionally, the black curves represent the Born-Huang corrected potentials $\W_1$ and $\W_2$.
In line with the color scheme used in figure \ref{fig:spec}, the eigenstates are obtained from the different models:
Born-Oppenheimer states in blue, Born-Huang states in orange, and exact states in green.
As the spectra suggest, most of the molecular wave functions obtained from the different models do not differ strongly. 
For example, the states that localize in the individual potential wells of the trilobite potential only deviate in energy.
These are the states that correspond to strong signals in figure \ref{fig:spec}.
Molecular states that have no counterpart in a different model localize in the vicinity of the avoided crossing where non-adiabatic couplings are particularly strong; these are located in the left-most region of each subfigure.

For $n=43$, in the left-most subfigure, most notably around $E=-3.68\,$GHz, a molecular state localizes at around $R=1650\,\mathrm{a}_0$. 
Although this state appears to be trapped in the Born-Huang potential well, the lack of a corresponding Born-Huang state shows that the potential well alone is not suitable to support such a state. 
In other words, this molecular state is trapped non-adiabatically by the off-diagonal derivative couplings.
The exact state has a very small but finite amplitude towards small $R$, resembling states trapped by internal quantum reflection, discussed extensively throughout the literature on ultralong-range Rydberg molecules \cite{Bendkowsky2010,Anderson2014th,Engel2019}.
At higher energy, around $E=-3.55\,$GHz, and at similar internuclear distance, there is a Born-Oppenheimer state localizing in a potential well that is shifted upwards by the Born-Huang correction.
This state cannot be identified in the spectrum because it is degenerate in energy with molecular states localizing at large internuclear distances, around $R=2500\,\mathrm{a}_0$, where non-adiabatic couplings are generally much weaker.
The corresponding Born-Huang state is at approximately $E=-3.53\,$GHz, barely visible in the figure.
There is no exact counterpart to these state.
In other words, these states are subject to rapid non-adiabatic decay into the continuum of the lower potential. 
At $n=44$, in the middle subfigure, the only region with significant differences between different models is again the potential formed by the avoided crossing, where single channel states localize, but exact states do not occur. 
At $n=45$, in the right-most subfigure, the overall structure is very similar to $n=43$.

To summarize the most important observations here, we find that the potential well formed by the avoided crossing of the trilobite and butterfly does not support localized molecular bound states as would be suggested by single channel calculations.
This is in marked contrast to other types of Rydberg molecules, most prominently Rydberg macrodimers, where a potential well is formed by the crossing of induced dipole potentials, and in charged Rydberg molecules, where a potential well is formed by the crossing of an attractive polarization potential and a repulsive dipole potential. 
Both of these cases are known to support bound molecular states \cite{Hollerith2019, Zuber2022}.
In ultralong-range Rydberg molecules, non-adiabatic couplings are too strong for such states to exist.
On the other hand, non-adiabatic couplings can lead to the formation of quantum reflection type states at the same radial position, but at energies much lower than the potential well.
Counterintuitively, both the observed effects, non-adiabatic decay and non-adiabatic trapping are more pronounced at principle quantum numbers where the derivative coupling element features a double peak structure.
In the case of a single narrow peak of $P_{12}$, such effects are less significant.
We attribute this to the fact that non-adiabatic couplings are strongly pronounced over a much wider range of $R$ in double peak situations.
Opposed to this, the single peak can become arbitrarily narrow.
Its integral converges at $\frac{\pi}{2}$ corresponding to a complete inversion of the adiabatic character.

Overall, the vibronic effects described in this section are not limited to the quantum numbers $n=43$ to 45.
On the contrary, we find similar patterns over a wide range of $n=25$ to 70.
Independent of $n$, the total energy shift between exact molecular states and states obtained within the Born-Oppenheimer approximation is approximately 5~MHz.
For states localizing in the vicinity of the avoided crossing, the shift can be much larger.
Vibronic interactions effects are less pronounced at principle quantum numbers featuring highly singular $P$-matrix elements resembling conical intersections.
These special situations occur repeatedly throughout the range of $n$ between 25 and 70.

\section{Conclusions \& Outlook \label{sec:conclusion}}

We have studied vibronic interaction effects in ultralong-range Rydberg molecules focusing on trilobite and butterfly states.
The avoided crossing between these two states gives rise to non-adiabatic trapping of molecular states including internal quantum reflection resonances.
Non-adiabatic decay is also observed.
Overall, non-adiabatic corrections to the adiabatic molecular structure lead to energy shifts of spectral lines on the MHz scale.
Such energy shifts may have been falsely attributed to corrections of electron phase shifts in the past.
Generally, in ultralong-range Rydberg molecules, a larger scattering lengths corresponds to deeper wells in the adiabatic potential energy curves.
This is why global energy shifts caused by diagonal non-adiabatic couplings can easily be mistaken for for a smaller scattering length or volume.

This work opens a pathway to study non-adiabatic interaction effects in Rydberg molecules.
We consider multiple scenarios to be of high relevance:
Firstly, non-adiabatic decay of molecular states correlated to Rydberg quantum defect states.
This is important to understand molecular lifetimes.
Secondly, non-adiabatic effects are naturally more pronounced in lighter molecules such as Li*Li, Li*Rb, or Na*Na, where they might lead to more striking effects.
Thirdly, the methods developed for this study can be applied to other species of Rydberg molecules such as charged molecules and macrodimers.
Lastly, exposing molecules to additional external fields is a promising perspective, especially as the gap between trilobite and butterfly curves can be controlled by an external electric field \cite{Kurz2013}. 
It may lead to more control of molecular properties, while causing more complexity, such as many-state interaction dynamics.

Notably, in the scenario considered here, the molecular channels are coupled but closed.
In ultralong-range Rydberg molecules, however, scenarios can be investigated involving open and closed channels. 
This introduces coupling to continuum states and can lead for example to Feshbach resonances.
In general, Rydberg molecules are highly asymmetric and imbalanced objects. 
Similar systems are encountered in semi-conductor materials based on Rydberg excitons.
Their non-adiabatic interaction effects may lead to unexpected material response properties.

\begin{acknowledgements}
	The authors gratefully acknowledge valuable discussions on diabatization strategies with Hossein R. Sadeghpour and Oriol Vendrell as well as discussions on numerical delicacies in the treatment of non-adiabatic couplings with Panagiotis Giannakeas.
	P. S. acknowledges support from the German Research Foundation (DFG) within the priority program "Giant Interactions in Rydberg Systems" [DFG SPP 1929 GiRyd project SCHM 885/30-2].
\end{acknowledgements}

\appendix
	
\section{Neglecting quantum defect states \label{sec:quantum-defects}}

\begin{figure}
	\centering
	\includegraphics[width=0.4\textwidth]{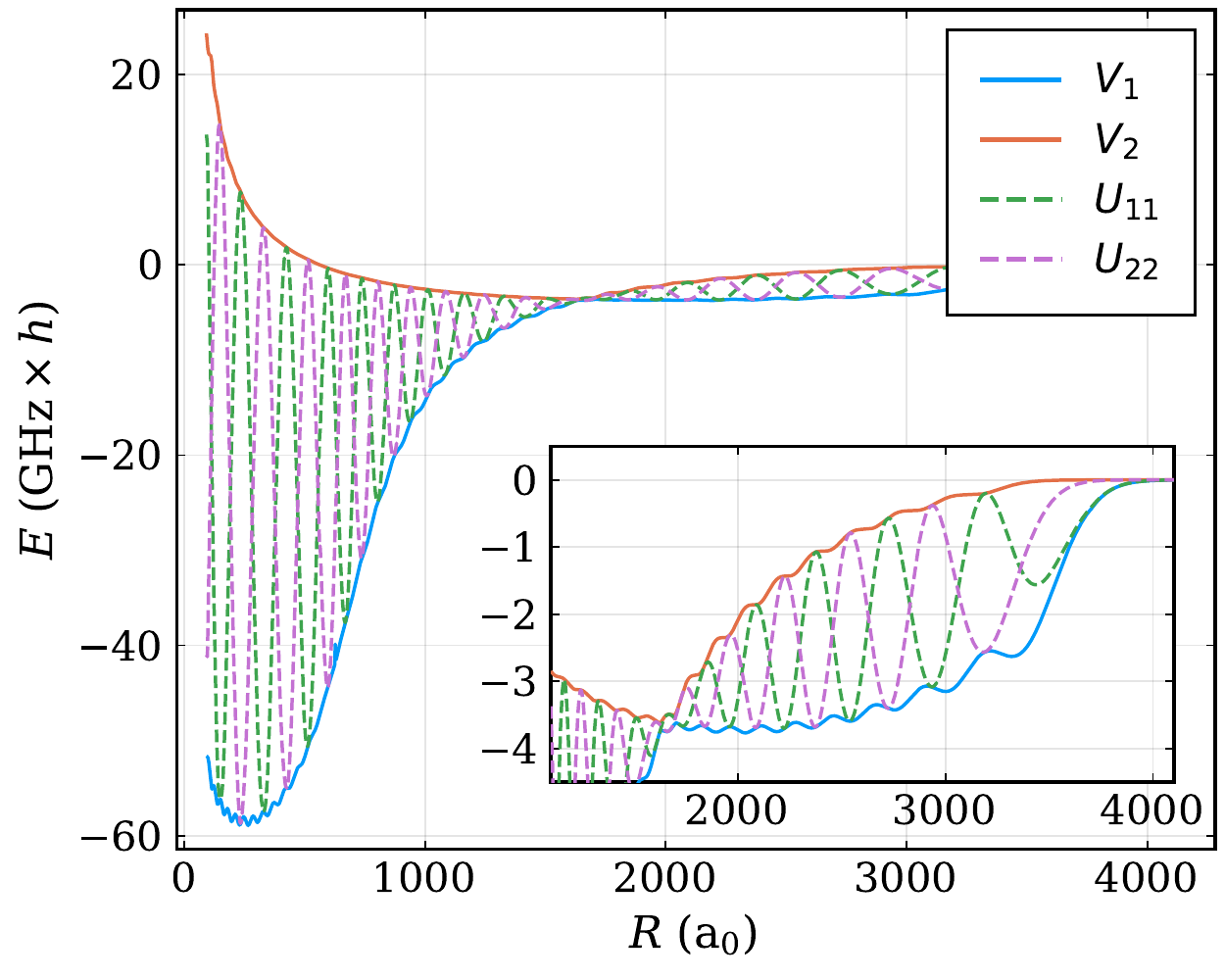}
	\caption[Adiabatic and diabatic potentials for $n=43$ neglecting quantum defects]{Adiabatic potential energy curves $\V_1$, $V_2$, and the diagonal diabatic potential matrix elements $\D_{11}$, $\D_{22}$ for principle quantum number $n=43$ neglecting quantum defect states. The inset shows a zoom. Energies are given relative to the degenerate atomic Rydberg level.}
	\label{fig:qd}
\end{figure}

To highlight the counterintuitive structure of the diabatic potential curves in the case of the trilobite and butterfly, figure \ref{fig:qd} shows potentials in an artificial setup.
Here, all quantum defects of rubidium have been set to zero.
This can also be interpreted as the electronic structure of H*Rb molecules.
Only two non-trivial adiabatic eigenstates remain:
The trilobite and the butterfly featuring only a single avoided crossing (solid curves).
The diabatic potentials (dotted curves) are oscillatory over the full range of $R$. 
The $P$-matrix element of these two states is large over the full range of $R$ leading to this behavior. 
This is related to the fact that the two states have an extremely large dipole transition element, orders of magnitude larger than for low-angular momentum states.

\section{Dipole diabatization \label{sec:dipole}}

Another way to represent the coupled system of equations (\ref{eq:SE}) is diabatization by aid of diagonalizing the electronic dipole transition matrix, i.e.~the position operator
\begin{equation}
	D_{\alpha\alpha'} = \int \mathrm{d} \vec{r} \, \vec{r} \psi^\ast_\alpha(\vec{r};R) \psi_{\alpha'}(\vec{r};R).
\end{equation}
The goal is to find a representation of the electronic basis that is dipole decoupled. 
Diagonalizing $\op{D}(R)$ equips us with a transformation $\bar{\vec{S}}(R)$ between the adiabatic and the dipole diabatic representation. 
It is defined via
\begin{equation}
	\op{d}= \bar{\vec{S}}^\dagger \op{D} \bar{\vec{S}},
\end{equation}
where $\op{d}(R)$ is a diagonal matrix. 
The Schrödinger equation becomes 
\begin{equation}
	\left[ \frac{-\nabla_R^2}{2\mu}+\bar{\vec{U}}-E\right] \bar{\vec{\chi}}=\frac{1}{2\mu}\bar{\vec{\Lambda}} \bar{\vec{\chi}}, 
\end{equation} 
where $\bar{\vec{\chi}}=\bar{\vec{S}}^\dagger \vec{\chi}$, $\bar{\vec{U}}=\bar{\vec{S}}^\dagger \op{\V} \bar{\vec{S}}$, and $\bar{\vec{\Lambda}}=\bar{\vec{S}}^\dagger \op{\Lambda} \bar{\vec{S}}$. 
While this is formally equivalent to equations (\ref{eq:SE}) and (\ref{eq:diaSE}), the point of this representation is that residual couplings $\bar{\vec{\Lambda}}$ are expected to be small, such that the right hand side of this equation is set to zero.
In the following, we will show that this is not an accurate assumption in our setup.

\section{Hybridized basis \label{sec:hybrid}}

\begin{figure*}
	\centering
	\includegraphics[width=\textwidth]{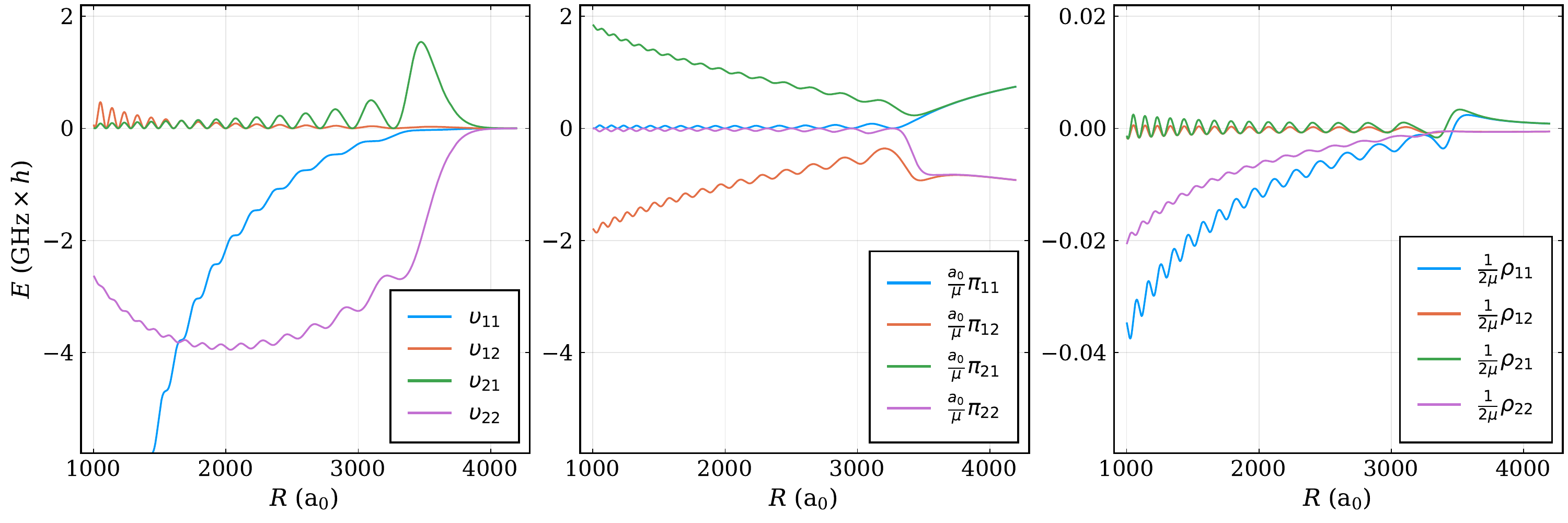}
	\caption[Potentials and couplings for $n=43$ in the hybridized basis]{Potential energy curves $\upsilon_{\kappa\kappa'}$ (left), derivative $\pi_{\kappa\kappa'}$ (middle), and scalar coupling elements $\rho_{\kappa\kappa'}$ (right) for principle quantum number $n=43$ obtained via the hybridized basis approach. Energies are given relative to the degenerate atomic Rydberg level.}
	\label{fig:hybpot}
\end{figure*}

The dipole diabatic representation is closely related to another approach originating in results found in references \cite{Chibisov2000, Cherkani2001} and developed previously in references \cite{Eiles2016, Eiles2019, Eiles2020}. 
Here, a hybridized electronic basis employing analytic functions of the trilobite and butterfly states 
\begin{align}
	\ket{1} &= \frac{1}{T} \sum_\beta\phi_\beta(R)\phi_\beta(\vec{r}), \\
	\ket{2} &= \frac{1}{B} \sum_\beta\phi'_\beta(R)\phi_\beta(\vec{r}),
\end{align}
with two ($R$-dependent) normalization constants $T^2=\sum_\beta|\phi_\beta|^2$ and $B^2=\sum_\beta|\phi'_\beta|^2$ is used to expand the total wavefunction. 
This ansatz suffers from the fact that the two basis states are not orthogonal. 
In fact, for large $R$, they become linearly dependent.
The off-diagonal overlap matrix element here is $\braket{1|2}=\frac{M^2}{TB}$, where $M^2=\sum\phi'(R)\phi(R)$. 
Still, the resulting potential- and non-adiabatic coupling terms provide some intuitive understanding of the physical situation. 
Employing the hybridized basis, the total Schrödinger equation contains a non-diagonal potential matrix
\begin{equation}
	\hat{\vec{\upsilon}} = \sum_{\kappa\kappa'} O^{-1}_{\kappa\kappa'} \braket{\kappa|V_{ea}|\kappa'} = 
	\begin{pmatrix}
		a_s T^2 & a_s \frac{T}{B} M^2 \\
		a_p \frac{B}{T} M^2 & a_p B^2
	\end{pmatrix},
\end{equation}
where $\kappa=\{1,2\}$ labels the two basis states, and $O_{\kappa\kappa'}^{-1}$ is the matrix element of the inverse overlap matrix.
The potential curves in $\hat{\vec{\upsilon}}$ almost exactly reproduce the dipole diabatic potentials for the trilobite and butterfly states in $\bar{\vec{U}}$. 
We attribute the slight differences merely to the fact that dipole diabatization provides an orthogonal basis. 
The elements of $\hat{\vec{\upsilon}}$ are shown in figure \ref{fig:hybpot} on the left.
Although this approach reduces the singularity of non-adiabatic couplings, they remain nonzero in this approach.
The derivative coupling matrix 
\begin{align}
	\nonumber \hat{\vec{\pi}} &= \sum_{\kappa\kappa'} O^{-1}_{\kappa\kappa'} \braket{\kappa|\nabla_R|\kappa'} \\
	&=
	\begin{pmatrix}
		-\frac{M^2(\tilde{M}^2-M^2N^2/B^2)}{T^2B^2-M^4} & \frac{B}{T} \\
		\frac{TB(\tilde{M}^2-M^2N^2/B^2)}{T^2B^2-M^4} & -\frac{M^2}{T^2}
	\end{pmatrix},
\end{align}
is not anti-hermitian, due the non-orthogonality of the underlying basis states.
The additional terms encountered in this expression are $\tilde{M}^2=\sum\phi''(R)\phi(R)$ and $N^2=\sum\phi''(R)\phi'(R)$.
For the same reason, the scalar coupling matrix $\hat{\vec{\rho}}$ does not obey the usual relation (\ref{eq:PQ}).
In fact, given this non-orthogonal basis, it is not clear how to analytically obtain a transformation to the adiabatic representation.
After all, this transformation is not unitary at large $R$.
While the analytic matrix elements of the potential and couplings are not straight forward, they all result in smooth functions of $R$ and $n$, which are shown in figure \ref{fig:hybpot}.
Notably, the derivative coupling elements (middle) become large asymptotically at large $R$.

\section{Adiabatic to diabatic transformations \label{sec:transformation}}

The unitary transformation between adiabatic and diabatic representation can be found analytically, as long as there is only a single adiabatic parameter (in our case $R$).
As these transformations are often useful, however, to our knowledge typically not provided in the literature, we reproduce here the explicit transformations for three and four channel systems. 
Considering three adiabatic states, the unitary transformation is a general rotation in three dimensions that can be expressed via three angles, e.g.~the Euler angles. 
Here, we choose
\begin{equation}
	\op{\U}_{3D}(R) = \op{R}_Z(\alpha) \cdot \op{R}_X(\pi/2-\beta) \cdot \op{R}_Z(\gamma),
\end{equation}
where $\op{R}_i(\varphi)$ is a rotation in three dimensions around the axis $i$ by the angle $\varphi$. 
Solving equation (\ref{eq:trans}) for a general three-state $P$ matrix
\begin{equation}
	\op{P}_{3D}=
	\begin{pmatrix}
		0 & p_{12} & p_{13} \\
		-p_{12} & 0 & p_{23} \\
		-p_{13} & -p_{23} & 0
	\end{pmatrix},
\end{equation}
leads to a set of coupled differential equations for the three angles $\alpha$, $\beta$, and $\gamma$:
\begin{align}
	\alpha' &= p_{12} - \tan\beta \, (p_{13} \cos\alpha + p_{23} \sin\alpha), \\
	\beta' &= p_{13} \sin\alpha - p_{23} \cos\alpha, \\
	\gamma' &= \sec\beta \, (p_{13} \cos\alpha + p_{23} \sin\alpha).
\end{align}
This approach extends to higher dimensional systems. 
A general rotation in $d$ dimensions can be characterized by $d(d-1)/2$ angles governing simple rotations that leave a $(d-2)$-dimensional subspace unchanged.

In four dimensions, general rotations are parameterized by six angles. The derivation and solution of the corresponding differential equations gets increasingly involved for more states. Let's call the four axes in four dimensions $\{w, x, y, z\}$, then simple rotations leave a plane unchanged. Say $\op{R}_{ij}(\theta)$ is a rotation of angle $\theta$ in the $i$-$j$-plane, e.g.
\begin{equation}
	\op{R}_{wy} = 
	\begin{pmatrix}
		\cos\theta & 0 & -\sin\theta & 0 \\
		0 & 1 & 0 & 0 \\
		\sin\theta & 0 & \cos\theta & 0 \\
		0 & 0 & 0 & 1
	\end{pmatrix}.
\end{equation}
We choose (somewhat arbitrary) an ansatz for a unitary matrix allowing general rotations in four dimensions as
\begin{align}
	\nonumber \op{\U}_{4D} = \ &\op{R}_{wx}(\alpha) \cdot \op{R}_{wy}(\beta) \cdot \op{R}_{wz}(\gamma) \\
	&\cdot \op{R}_{xy}(-\delta) \cdot \op{R}_{xz}(\epsilon) \cdot \op{R}_{yz}(\zeta)
\end{align}
and the general $P$ matrix
\begin{equation}
	\op{P}_{4D}=
	\begin{pmatrix}
		0 & p_{12} & p_{13} & p_{14} \\
		-p_{12} & 0 & p_{23} & p_{24} \\
		-p_{13} & -p_{23} & 0 & p_{34} \\
		-p_{14} & -p_{24} & -p_{34} & 0
	\end{pmatrix}.
\end{equation}
The angle $\alpha$ mixes $P$-matrix elements leading to the following short-hand notations
\begin{align}
	a_3^- &= p_{13} \sin\alpha - p_{23} \cos\alpha, \\
	a_4^- &= p_{14} \sin\alpha - p_{24} \cos\alpha, \\
	a_3^+ &= p_{13} \cos\alpha + p_{23} \sin\alpha, \\
	a_4^+ &= p_{14} \cos\alpha + p_{24} \sin\alpha,
\end{align}
and like-wise for $\beta$ 
\begin{align}
	b^- &= a_4^+ \sin\beta - p_{34} \cos\beta, \\
	b^+ &= a_4^+ \cos\beta + p_{34} \sin\beta,
\end{align}
and like-wise for $\delta$ 
\begin{align}
	d^- &=  b^- \sin\delta - a_4^- \cos\delta, \\
	d^+ &= b^- \cos\delta + a_4^- \sin\delta,
\end{align}
such that the following set of differential equations is obtained for the six angles
\begin{align}
	\alpha' &= p_{12} + a_3^- \tan\beta + a_4^- \sec\beta \tan\gamma, \\
	\beta' &= a_3^+ + b^- \tan\gamma, \\
	\gamma' &= b^+, \\
	\delta' &= a_3^- \sec\beta + a_4^- \tan\beta \tan\gamma - d^- \sec\gamma \tan\epsilon, \\
	\epsilon' &= -d^+ \sec\gamma, \\
	\zeta' &= d^- \sec\gamma \sec\epsilon.
\end{align}

\bibliographystyle{apsrev4-2}
\bibliography{main.bbl}

\end{document}